\documentclass[conference]{IEEEtran}
\IEEEoverridecommandlockouts  
\usepackage{cite}

\usepackage{placeins}

\usepackage{soul}

\usepackage{float}

\usepackage{url}

\ifCLASSINFOpdf
\else
\fi
\usepackage{amsmath}
\usepackage{graphicx}
\usepackage{subcaption}

\usepackage{url}
\usepackage{graphicx} 

\usepackage{amssymb}

\usepackage[T1]{fontenc}
\usepackage{lmodern}
\usepackage{float} 
\usepackage{titlesec} 

\titlespacing{\section}{0pt}{5pt}{3pt}
\titlespacing{\subsection}{0pt}{4pt}{2pt}

\begin{document}
%
\title{Impact of Altitude, Bandwidth, and NLOS Bias on TDOA-Based 3D UAV Localization: Experimental Results and CRLB Analysis}

\author{   
    \IEEEauthorblockN{Cole Dickerson\IEEEauthorrefmark{1}, Saad Masrur\IEEEauthorrefmark{1}, Jonah Dickerson\IEEEauthorrefmark{2},\"{O}zg\"{u}r \"{O}zdemir
\IEEEauthorrefmark{1}, Ismail G\"{u}ven\c{c}
\IEEEauthorrefmark{1}
    }
    \IEEEauthorblockA{\IEEEauthorrefmark{1}Dept. of Electrical and Computer Engineering, North Carolina State University, Raleigh, USA}
    \IEEEauthorblockA{\IEEEauthorrefmark{2}Dept. of Geography, Planning, and Environment, East Carolina University, Greenville, USA}
    \IEEEauthorblockA{Email: \{jcdicker, smasrur, oozdemi, iguvenc\}@ncsu.edu, dickersonj24@students.ecu.edu}  

}


%


\maketitle
\begin{abstract}
This paper investigates unmanned aerial vehicle (UAV) localization using time difference of arrival (TDOA) measurements under mixed line-of-sight (LOS) and non-line-of-sight (NLOS) conditions. A 3D TDOA Cramér-Rao lower bound (CRLB) model is developed accounting for varying altitudes and signal bandwidths. The model is compared to five real-world UAV flight experiments conducted at different altitudes (40 m, 70 m, 100 m) and bandwidths (1.25 MHz, 2.5 MHz, 5 MHz) using Keysight N6841A radio frequency (RF) sensors of the NSF AERPAW platform. Results show that altitude, bandwidth, and NLOS obstructions significantly impact localization accuracy. Higher bandwidths enhance signal time resolution, while increased altitudes mitigate multipath and NLOS biases, both contributing to improved performance. However, hovering close to RF sensors degrades accuracy due to antenna pattern misalignment and geometric dilution of precision. These findings emphasize the inadequacy of traditional LOS-based models in NLOS environments and highlight the importance of adaptive approaches for accurate localization in challenging scenarios.
\end{abstract}


%
\IEEEpeerreviewmaketitle

\section{Introduction}
Unmanned aerial vehicles (UAVs) are increasingly popular due to their applications in delivery, agriculture, aerial base stations, search-and-rescue, and wireless spectrum monitoring and enforcement. The growing number of UAVs necessitates a reliable air-traffic management system to prevent congestion. Governments, industries, and manufacturers are collaborating to develop a secure UAV traffic management (UTM) system \cite{faa2020utm} where accurate localization is vital to ensure safe separation and collision prevention. Recent studies have explored using radio frequency (RF) signals for UAV detection, classification, localization, and tracking, even for non-cooperative or malicious drones \cite{guvenc2018detection, azari2018key, boon2017rf}. Among these techniques, Time Difference of Arrival (TDOA), a multilateration-based method that localizes UAVs by measuring signal arrival time differences across spatially separated sensors, has proven particularly effective for passive RF sensing \cite{boon2017rf}.

Numerous studies have explored TDOA-based localization for UAVs due to its effectiveness in passive RF sensing. \cite{sinha2019impact} examines the impact of 3D antenna radiation patterns on TDOA accuracy, deriving the Cram\'er-Rao lower bound (CRLB) and evaluating performance through analytical modeling and simulations. \cite{bhattacherjee2022experimental} presents an experimental study on UAV localization using Keysight RF sensors, proposing a motion model-based extended Kalman filter to improve tracking in GPS-denied environments. \cite{8115937} introduces a scalable TDOA ultra-wideband system for multi-UAV indoor navigation, leveraging wireless clock synchronization to enhance accuracy. \cite{khan2025cramer} investigates Time of Arrival (TOA) localization with directive antennas, deriving new Fisher Information Matrix (FIM) expressions and highlighting improved feasibility in challenging scenarios. 

Building on these studies, this paper introduces a CRLB model for UAV localization that incorporates 3D coordinates, varying signal bandwidths, and signal-to-noise ratios (SNRs). The model is compared to data from five real-world UAV flights at different altitudes and bandwidths, tracked using Keysight N6841A RF sensors. Experiments were conducted at the Aerial Experimentation and Research Platform for Advanced Wireless (AERPAW), a state-of-the-art testbed designed for UAV and 5G integration, which provided the infrastructure for precise, repeatable flight trajectories and RF data collection. To the best of our knowledge, this is the first study to experimentally study a CRLB model for 3D UAV localization under such conditions. Our key contributions are: 1) an in-depth analysis of how signal bandwidth and UAV altitude affect localization accuracy; 2) a comparison of the CRLB to real-world UAV localization data, assessing the alignment between theoretical predictions and experimental results; and 3) the release of a dataset containing UAV positioning and LOS/NLOS data to support future research, publicly available at \cite{AERPAW2024}.

\section{System Model}

\subsection{Derivation of TOA Measurements from RF Signals}
The TDOA measurement for a pair of time-synchronized RF sensors is obtained by subtracting the TOA estimates at each sensor, eliminating dependency on the UAV's unknown transmission time. Using a maximum-likelihood estimator (ML) for TOA ensures the measurement noise can be approximated as zero-mean Gaussian due to the asymptotic normality of ML estimators \cite{seyed2022handbook}. In a network of $N$ passive RF sensors or fixed terminals (FTs), with a UAV as the transmitter and the RF sensors as receivers, the UAV's true location is $\mathbf{x} = [x,y,z]^T$, its estimated location is $\hat{\mathbf{x}} = [\hat{x},\hat{y},\hat{z}]^T$ 
and the position of the $i$th FT is given by 
$\mathbf{x}_i = [x_i,y_i,z_i]^T$. The measured distance between the UAV and the $i$th FT, $\hat{d}_i$, can be modeled as
\begin{equation}
\hat{d}_i=d_i+b_i+n_i=c t_i, \quad i=1,2, \ldots, N,
\end{equation}

\noindent where $d_i$ is the real distance between the UAV and the $i$th FT, $n_i \sim \mathcal{N}\left(0, \sigma_i^2\right)$ is additive white Gaussian noise (AWGN) with variance $\sigma_i^2$, $c$ is the speed of light, $t_i$ is the TOA of the signal at the $i$th FT, and $b_i$ is a positive distance bias introduced due to the blockage of line of sight transmission paths which has been modeled as Gaussian \cite{sinha2022impact}, exponentially \cite{gezici2004uwb}, and uniformly \cite{venkatesh2006nlos} distributed in the literature. In this work, we derive a CRLB assuming a LOS-dominated environment. While NLOS conditions introduce additional biases that impact localization accuracy, explicitly modeling these effects requires site-specific parameterization and increases mathematical complexity. As a first step, we focus on establishing a LOS-based benchmark for TDOA UAV localization representing the best possible performance, with future work incorporating NLOS effects through empirical modeling. Assuming zero NLOS bias, the ML estimator achieves asymptotic optimality in a LOS-dominated environment, with the TOA estimate covariance reaching the CRLB at all sensors, represented as the inverse FIM, $\mathbf{I}(\tau)$. Consequently, the TOA measurement noise at the $i$-th sensor in equation (1) is modeled as zero-mean Gaussian with a variance of $\sigma_i^2= \left[\mathbf{I}^{-1}(\tau)\right]_{i i}$, expressed as \cite{gezici2008survey}:
\begin{equation}
\sigma_i=\frac{1}{2 \sqrt{2} \pi \sqrt{\text{SNR}_i} \beta},
\label{eq:toa_crlb}
\end{equation}
\noindent where $\beta$ is the effective UAV signal bandwidth, and $\text{SNR}_i = P_\text{r,i} / P_\text{n}$. The received power $P_\text{r,i}$ at sensor $i$ is commonly modeled using the Free Space Path Loss (FSPL) formula:
\begin{equation}
P_\text{r,i}=P_\text{t} G_\text{t} G_\text{r,i} \frac{\lambda^2}{(4 \pi d_i)^2}.
\label{eq:receivedpower}
\end{equation}
In (\ref{eq:receivedpower}), $P_\text{t}$ is the transmitted power, $G_\text{t}$ and $G_\text{r,i}$ are the gains of the transmitting and $i$-th receiving antennas, $\lambda$ is the signal wavelength, and $d_i$ is the distance between the transmitter and sensor $i$. The thermal noise power $P_\text{n}$ is modeled as $P_\text{n}=k T B$, where $k$ is Boltzmann's constant ($1.38 \text{x} 10^{-23} \text{J/K}$), $T$ is the absolute temperature in Kelvin, and $B$ is the bandwidth of the communication system. 
\begin{figure}
    \centering
    \includegraphics[width=1\linewidth]{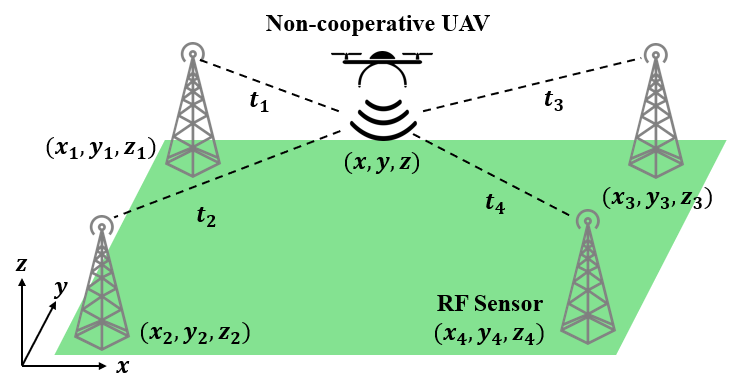}
    \caption{System model for passive localization of a non-cooperative UAV in 3D with four FTs.}
    \label{fig:TOAsysmodel}
\end{figure}
Fig.~\ref{fig:TOAsysmodel} illustrates a system with four RF sensors/FTs located at $\mathbf{x}_i = [x_i,y_i,z_i]^T$, $i = 1,...,4$, and a non-cooperative UAV at $\mathbf{x} = [x,y,z]^T$, with TOA measurements $(t_1,...,t_4)$ for the signals received at each RF sensor.

\subsection{Computation of TDOA from TOA Estimates}
TDOA estimation, which does not require clock synchronization between the UAV and FTs, is ideal for non-cooperative sources but requires synchronization among the FTs. TDOA calculates the difference in signal arrival times at two FTs, placing the UAV on a hyperbola with the FTs as foci \cite{gezici2008survey}. For 3D localization, four sensors are needed to determine the UAV's position via the intersection of three hyperboloids \cite{seyed2022handbook}. One FT is chosen as the reference, and TDOA measurements for the remaining $N-1$ sensors are computed relative to it. The TDOA measurement $\Delta t_{r,i}$
for FT $i$ using FT $r$ as the reference is given by:
\begin{equation*}
\Delta t_{r,i} = t_r - t_i = \frac{\hat{d}_r - \hat{d}_i}{c} = \frac{\left(d_r + b_r + n_r\right) - \left(d_i + b_i + n_i\right)}{c},
\end{equation*}
\noindent where $r, i \in \{1, \ldots, N\}, \ r \neq i$, and $r$ is fixed. This can be reorganized to separate the deterministic part (based on true distances $d_r$ and $d_i$) from stochastic noise and NLOS bias:
\begin{equation}
\Delta t_{r,i}=h_{r,i}(\mathbf{x})+v_{r,i},
\end{equation}
\noindent where $h_{r,i}(\mathbf{x})=\frac{d_r(\mathbf{x})-d_i(\mathbf{x})}{c}$
is the range difference, which depends on the position $\mathbf{x}$ of the UAV, and $v_{r,i}=\frac{\left(n_r-n_i\right)+\left(b_r-b_i\right)}{c}$ which represents the combined measurement noise and NLOS bias for the pair $(r,i)$. If we continue to assume NLOS bias = 0, $\Delta t_{r,i}$ is a Gaussian random variable with mean and variance:
\begin{equation}
\Delta t_{r,i} \sim \mathcal{N}\left(\frac{d_r-d_i}{c}, \sigma_r^2+\sigma_i^2\right).
\end{equation}
Since TDOA measurements share a common reference sensor $r$, they are not independent, as reflected in the covariance matrix $\mathbf{Q}(\mathbf{x})$. The TDOA measurement vector $\Delta t = [\Delta t_{r,i}, \Delta t_{r,(i+1)}, ... , \Delta t_{r,N}]^T$ follows a joint Gaussian distribution, $\Delta t \sim\mathcal{N}(\mu(\mathbf{x}), \mathbf{Q}(\mathbf{x}))$, where the mean vector and covariance matrix depend on the UAV's unknown position, $\mathbf{x}$ and are given by \cite{sinha2019impact}:
\begin{align}
\boldsymbol{\mu}(\mathbf{x}) &= \frac{1}{c} \begin{bmatrix}
d_r - d_i, d_r - d_{i+1}, \ldots, d_r - d_N
\end{bmatrix}^T, \\
\mathbf{Q}(\mathbf{x}) &= \begin{pmatrix}
\sigma_r^2 + \sigma_i^2 & \sigma_r^2 & \cdots & \sigma_r^2 \\
\sigma_r^2 & \sigma_r^2 + \sigma_{i+1}^2 & \cdots & \sigma_r^2 \\
\vdots & \vdots & \ddots & \vdots \\
\sigma_r^2 & \sigma_r^2 & \cdots & \sigma_r^2 + \sigma_N^2
\end{pmatrix}.
\end{align}
Additionally, $\mathbf{s}=\left[s_1, s_2, \ldots, s_N\right]$ is the LOS/NLOS indicator vector for $N$ sensors, where each element $s_i$ is:
\begin{equation}
s_i= \begin{cases}1, & \text { if there is a LOS path to sensor } i \\ 0, & \text { if there is an NLOS path to sensor } i\end{cases}.
\end{equation} 
The LOS/NLOS indicator vector$\mathbf{s}$ is excluded from the CRLB results but is used in the ray-tracing simulations.

\section{LOS CRLB for TDOA-Based UAV Localization}

\begin{figure*}[ht] 
    \centering
    \begin{subfigure}[b]{0.3\textwidth}
        \centering
        \includegraphics[height=5cm,keepaspectratio]{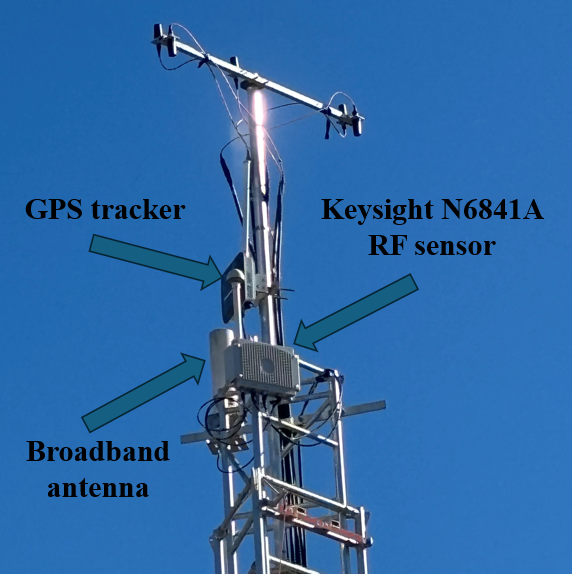}
        \caption{Keysight N6841A RF sensor deployed on AERPAW tower LW3.}
        \label{RFsensor}
    \end{subfigure}
    \hfill
    \begin{subfigure}[b]{0.3\textwidth}
        \centering
        \includegraphics[height=5cm,keepaspectratio]{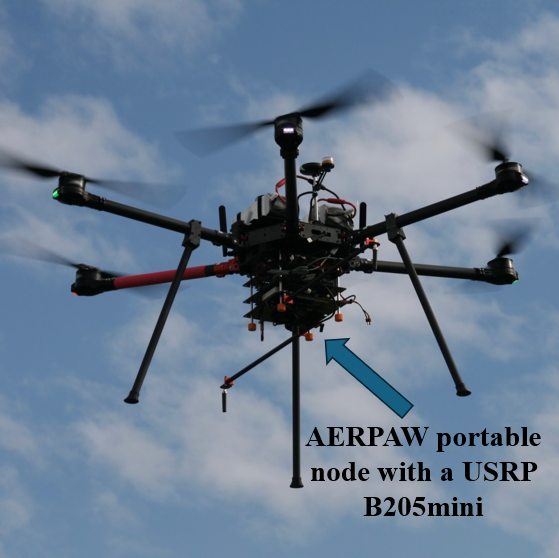}
        \caption{AERPAW UAV with SDR portable node.}
        \label{UAV}
    \end{subfigure}
    \hfill
    \begin{subfigure}[b]{0.3\textwidth}
        \centering
        \includegraphics[height=5cm,keepaspectratio]{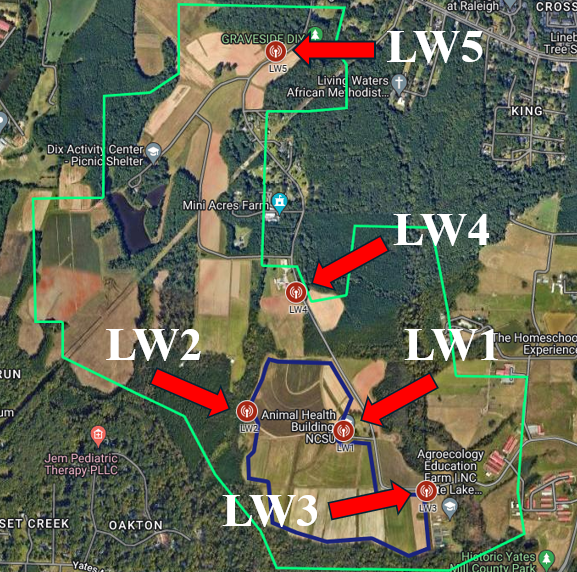}
        \caption{AERPAW's LWFRL and RF sensor tower locations.}
        \label{LWRFL}
    \end{subfigure}

    \caption{Overview of the AERPAW testbed showing the RF sensor, UAV node, and Lake Wheeler Field Road site layout.}
    \label{fig:aerpaw_overview}
\end{figure*}

The CRLB represents the theoretical minimum variance achievable by any unbiased estimator, serving as a benchmark for evaluating UAV localization accuracy. It is derived from the inverse of the FIM, $\mathbf{I}_{\hat{\mathbf{x}}}$, which quantifies the information content of measurements about the UAV's position $\mathbf{x} = (x,y,z)$. For TDOA-based UAV localization, the CRLB bounds the error covariance of the position estimate $\mathbf{\hat{x}}$ derived from the measurement vector $\Delta t$. This relationship is formally expressed as:
\begin{equation}
\mathbb{E}\left[(\hat{\mathbf{x}}-\mathbf{x})(\hat{\mathbf{x}}-\mathbf{x})^T\right] \geq \mathbf{I}_{\hat{\mathbf{x}}}^{-1},
\end{equation}
\noindent where $\mathbb{E}[\cdot]$ is the expectation operator. The FIM for TDOA localization at a single time step is given by \cite{seyed2022handbook}:
\begin{equation}
\mathbf{I}_{\hat{\mathbf{x}}} = \frac{\partial \Delta t^\mathrm{T}}{\partial \mathbf{x}} \mathbf{Q}^{-1}(\mathbf{x}) \frac{\partial \Delta t}{\partial \mathbf{x}}.
\end{equation}
\noindent Here, $\mathbf{Q}(\mathbf{x})$ represents the covariance matrix of the  TDOA measurements, and $\frac{\partial \Delta t^\mathrm{T}}{\partial \mathbf{x}}$ is the Jacobian matrix of the measurement vector, $\Delta t$, expressed here as: 
\begingroup
\setlength\arraycolsep{2pt} 
\renewcommand{\arraystretch}{0.9} 
\begin{align*}
\scalebox{0.85}{$\frac{\partial \Delta t^\mathrm{T}}{\partial \mathbf{x}}$} &=
\begin{bmatrix}
\frac{\partial \Delta t_{r,i}}{\partial x} & \frac{\partial \Delta t_{r,i}}{\partial y} & \frac{\partial \Delta t_{r,i}}{\partial z} \\ 
\frac{\partial \Delta t_{r,(i+1)}}{\partial x} & \frac{\partial \Delta t_{r,(i+1)}}{\partial y} & \frac{\partial \Delta t_{r,(i+1)}}{\partial z} \\ 
\vdots & \vdots & \vdots \\ 
\frac{\partial \Delta t_{r,N}}{\partial x} & \frac{\partial \Delta t_{r,N}}{\partial y} & \frac{\partial \Delta t_{r,N}}{\partial z}
\end{bmatrix}, \\[1em]
&=
\begin{bmatrix}
\frac{x - x_r}{\ell_r} - \frac{x - x_i}{\ell_i} & \frac{y - y_r}{\ell_r} - \frac{y - y_i}{\ell_i} & \frac{z - z_r}{\ell_r} - \frac{z - z_i}{\ell_i} \\
\frac{x - x_r}{\ell_r} - \frac{x - x_{i+1}}{\ell_{i+1}} & \frac{y - y_r}{\ell_r} - \frac{y - y_{i+1}}{\ell_{i+1}} & \frac{z - z_r}{\ell_r} - \frac{z - z_{i+1}}{\ell_{i+1}} \\
\vdots & \vdots & \vdots \\
\frac{x - x_r}{\ell_r} - \frac{x - x_N}{\ell_N} & \frac{y - y_r}{\ell_r} - \frac{y - y_N}{\ell_N} & \frac{z - z_r}{\ell_r} - \frac{z - z_N}{\ell_N}
\end{bmatrix},
\end{align*}
\endgroup
\noindent where $\ell_{i}$ is the Euclidean distance between the UAV and sensor $i$:
\begin{equation*}
\ell_i = \|\mathbf{x} - \mathbf{x}^{(i)}\| = \sqrt{(x - x^{(i)})^2 + (y - y^{(i)})^2 + (z - z^{(i)})^2},
\end{equation*}
\noindent and where $i = 1, \ldots, N$. The FIM will be a 3x3 matrix of the form:
\begin{equation}
\mathbf{I}_{\hat{\boldsymbol{x}}}=\left(\begin{array}{lll}
F_{11} & F_{12} & F_{13} \\
F_{21} & F_{22} & F_{23} \\
F_{31} & F_{32} & F_{33}
\end{array}\right).
\end{equation}
\noindent $F_{11}$, $F_{22}$, and $F_{33}$ represent the Fisher Information for the $x$, $y$, and $z$ coordinates, respectively, while off-diagonal terms capture cross-information. The CRLB for the localization estimate is obtained by inverting the FIM and taking its trace: 
\begin{equation}
\text{CRLB} = \text{Tr}\left(\mathbf{I}_{\hat{\boldsymbol{x}}}^{-1}\right).
\end{equation}
\noindent The square root of the CRLB provides the Root Mean Square Error (RMSE), which quantifies the average localization error for the UAV in the same units as the estimated parameters:
\begin{equation}
\operatorname{RMSE}(x, y, h) \geq \sqrt{\operatorname{trace}[\operatorname{CRLB}(x, y, h)]}.
\end{equation}
The LOS CRLB provides a bound for localization accuracy, but its practical implications in mixed LOS/NLOS environments must be validated. The following section details the experimental setup and ray-tracing simulations used to identify LOS and NLOS paths, forming the basis for real-world comparison of the theoretical results.  

\section{Experimental Setup and Ray-Tracing Simulations}

\begin{figure*}[!h] 
    \centering
    \begin{subfigure}[b]{0.3\textwidth} 
        \centering 
        \includegraphics[width=\textwidth, keepaspectratio]{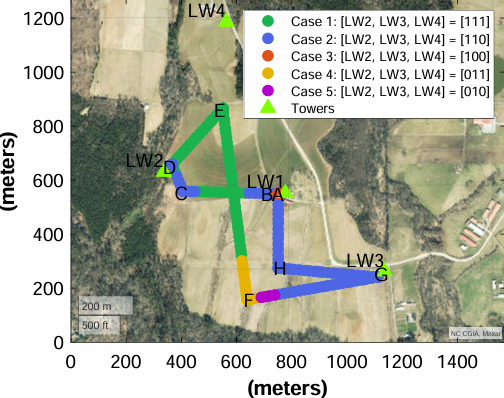} 
        \caption{40 m altitude.}
        \label{LOSat40m}
    \end{subfigure}
    \hfill
    \begin{subfigure}[b]{0.3\textwidth} 
        \centering 
        \includegraphics[width=\textwidth, keepaspectratio]{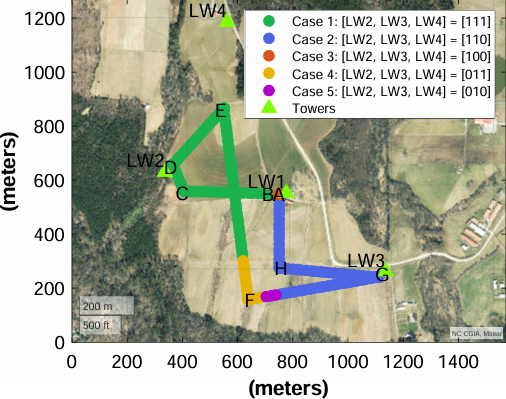} 
        \caption{70 m altitude.}
        \label{LOSat70m}
    \end{subfigure}
    \hfill
    \begin{subfigure}[b]{0.3\textwidth} 
        \centering 
        \includegraphics[width=\textwidth, keepaspectratio]{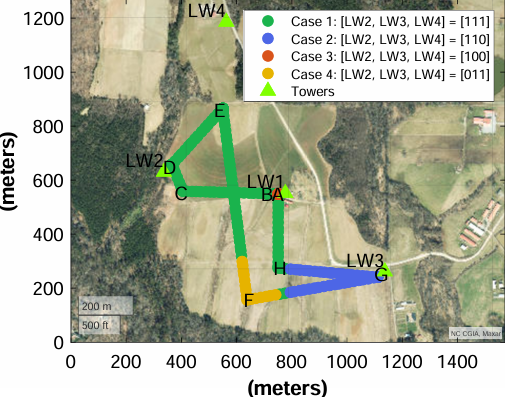} 
        \caption{100 m altitude.}
        \label{LOSat100m}
    \end{subfigure}
    \caption{LOS conditions at 40, 70, and 100 meter altitudes.}
    \label{fig:LOSconditions}
\end{figure*}

\begin{table*}[ht] 
\setcounter{table}{1}  
\captionsetup{aboveskip=2pt}  
\centering
\caption{Impact of altitude and bandwidth on TDOA-based 3D UAV localization.}
\label{errorresults}
\begin{tabular}{|l|c|c|c|}
\hline
\textbf{Altitude, Bandwidth} & \textbf{Valid Measurements (Count, \%)} & \textbf{Total Average Error} & \textbf{Average Error (Outliers Removed)} \\ \hline
40 m, 1.25 MHz                & 186 out of 255 (72.9\%)                  & 312.02 m                        & 79.06 m                         \\ \hline
40 m, 2.5 MHz                 & 168 out of 242 (69.4\%)                  & 280.46 m                        & 55.94 m                          \\ \hline
40 m, 5 MHz                   & 185 out of 245 (75.5\%)                  & 358.72 m                        & 48.88 m                           \\ \hline
70 m, 5 MHz                   & 205 out of 245 (83.7\%)                  & 256.55 m                       & 46.87 m                          \\ \hline
100 m, 5 MHz                  & 240 out of 261 (92.0\%)                  & 163.70 m                       & 35.84 m                          \\ \hline
\end{tabular}
\end{table*}

AERPAW is a state-of-the-art testbed for research at the convergence of 5G technology and autonomous UAVs. This study leveraged AERPAW's controlled environment to conduct UAV flight experiments and collect TDOA measurements using Keysight N6841A RF sensors under mixed LOS/NLOS scenarios. The N6841A (Fig.~\ref{RFsensor}), detects, records, and time-stamps RF signals with high precision, operating within a frequency range of 20 MHz to 6 GHz with up to 20 MHz bandwidth. It features a broadband omnidirectional antenna, GPS-based timestamping, and geolocation software (N6854A) that supports TDOA, received signal strength (RSS), and hybrid localization algorithms for RF source tracking within a 2 km radius. For UAV localization, the N6841A captures IQ components of RF signals, which are processed to estimate the UAV's position. 

AERPAW's Lake Wheeler Field Road Labs (LWFRL) host four Keysight RF sensors on towers LW2 through LW5 (Fig.~\ref{LWRFL}, red circles). Five UAV flights were conducted to evaluate localization performance under varying conditions. During each flight, a 3.32 GHz channel sounding waveform was transmitted via GNU Radio on the UAV's software defined radio (SDR) portable node (Fig.~\ref{UAV}), and RF sensors localized the UAV along its trajectory. Flight details, including altitude and bandwidth, are summarized in TABLE~\ref{table:flight_details}. Post-processing of estimated and ground-truth UAV positions was done in MATLAB. Noise variances, derived using ($\ref{eq:toa_crlb}$) and ($\ref{eq:receivedpower}$), depend on frequency, bandwidth, and UAV-sensor distance. The center frequency was 3.32 GHz, with effective bandwidths of 1.25, 2.5, or 5 MHz depending on the flight. The transmitter power was 30.67 dBm, and the average temperature was 304.3 K ($88^{\circ} \mathrm{F}$). Transmitter and receiver gains were assumed negligible for simplicity. 

\begin{table}[H]
\setcounter{table}{0}  
\caption{Flight variable summary.}
\centering
\resizebox{\columnwidth}{!}{  
    \begin{tabular}{|c|c|c|c|c|c|}
    \hline
    \textbf{Variable}       & \textbf{Flight 1} & \textbf{Flight 2} & \textbf{Flight 3} & \textbf{Flight 4} & \textbf{Flight 5} \\ \hline
    Altitude           & 40 m  & 40 m  & 40 m & 70 m & 100 m\\ \hline
    Bandwidth        & 1.25 MHz & 2.5 MHz & 5.0 MHz & 5.0 MHz& 5.0 MHz \\ \hline
    \end{tabular}
}
\label{table:flight_details}
\end{table}

\subsection{Ray-Tracing Simulations for LOS/NLOS Path Identification}

To identify areas where the LOS CRLB may not accurately characterize the lower bound on variance in a mixed LOS/NLOS environment, LOS and NLOS paths to each RF sensor along the UAV trajectories were analyzed. Environmental features like trees and buildings were modeled using JavaOpenStreetMap, then exported to MATLAB for ray-tracing simulations. These simulations accounted for reflection, diffraction, and obstructions to classify LOS and NLOS conditions along the UAV's trajectory.

Figs.~\ref{LOSat40m}, \ref{LOSat70m}, and \ref{LOSat100m} show the geographical layout and LOS conditions at each tower for UAV flights at 40, 70, and 100 m altitude, respectively. The flight path, overlaid on a satellite map, includes waypoints (A, B, C, etc.) and color-coded segments indicating LOS conditions (green: predominately LOS (3/4); blue and yellow: half LOS (2/4); red and purple (1/4): mostly NLOS) for sensors LW2, LW3, and LW4. LW5, excluded from the figures, but included in all TDOA measurements and CRLB analysis, experiences constant NLOS conditions due to obstructions. The map highlights environmental influences, such as open fields and wooded areas, on signal propagation and LOS status. 

\section{Experimental and Numerical Results}

This section presents the results of the experimental flights, divided into subsections that explore the effects of altitude and bandwidth on TDOA-based 3D UAV localization. TABLE II provides a summary of the number of valid measurements, total average localization error, and average localization error when outliers with RMSE larger than 200 m are removed, offering an overview of localization performance under varying conditions.

\subsection{Impact of Altitude on Localization Performance}

\begin{figure}[t]
    \centering
    \includegraphics[width=0.75\linewidth]{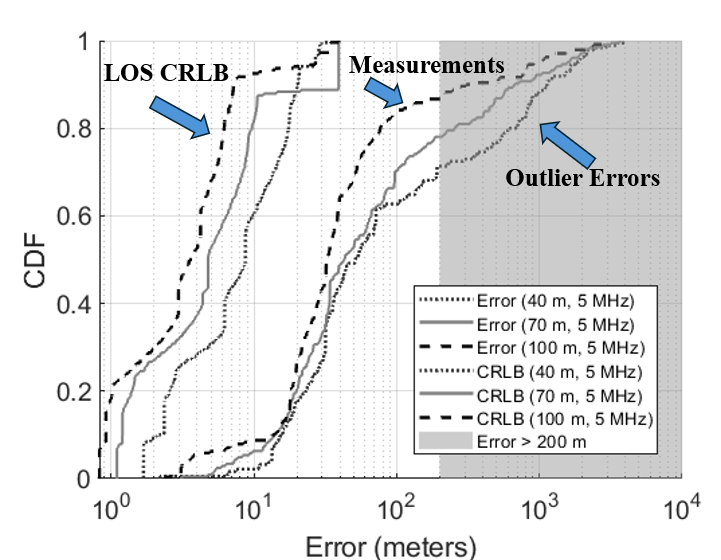}
    \caption{CDFs of localization errors and CRLBs for flights at 40 m, 70 m, and 100 m altitudes with a constant bandwidth of 5 MHz.}
    \label{fig:altcdfs}
\end{figure}

As shown in TABLE \ref{errorresults}, the total average error decreases significantly with each 30-meter increase in altitude, from 358.72 m to 256.55 m, and then to 163.70 m. Similarly, the average error when outliers are removed (error greater than 200 meters) decreases from 48.88 m to 46.87 m, and finally to 35.84 m. Fig.~\ref{fig:altcdfs} shows the cumulative distribution functions (CDFs) of errors and CRLBs for flights at varying altitudes with constant bandwidth, highlighting errors exceeding 200 m where a Kalman filter, as in \cite{bhattacherjee2022experimental}, could mitigate outliers. The experimental data demonstrates consistent error reduction with increasing altitude, and CRLBs closely follow the error trend until the upper 15th percentile. We note here that the CRLB results do not explicitly take into account NLOS bias effects, which we leave as future work. 

Improved localization accuracy at higher altitudes is due to several factors. Higher altitudes increase the likelihood of LOS with RF sensors, reducing NLOS biases from obstructions like buildings and trees. They also minimize multipath by increasing the distance from reflective ground surfaces, resulting in cleaner signal propagation. Additionally, altitude reduces disparities in distances to RF sensors, leading to more consistent SNR and RSS across sensors. Subtle differences in propagation dynamics and flight path geometry may also contribute to performance variations. 

\subsection{Impact of Bandwidth on Localization Performance}

\begin{figure}[t]
    \centering
    \includegraphics[width=0.75\linewidth]{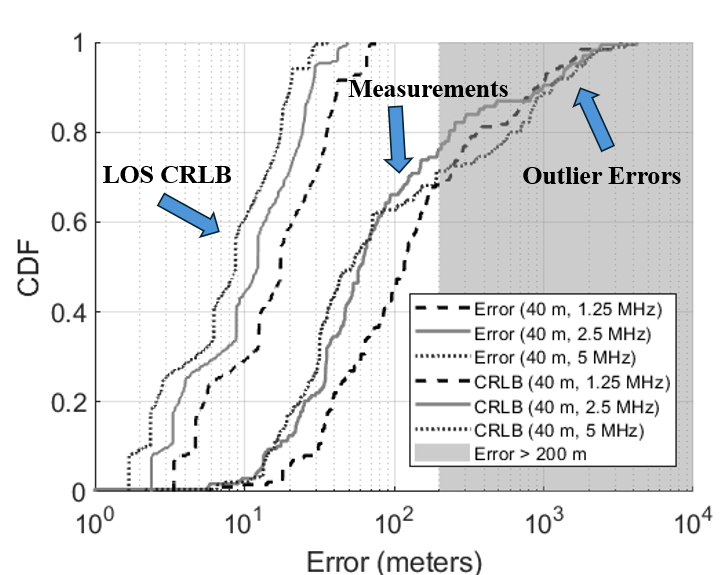}
    \caption{CDFs of localization errors and CRLBs for flights at 1.25 MHz, 2.5 MHz, and 5 MHz bandwidths with a constant altitude of 40 m.}
    \label{fig:bandwidthcdfs}
\end{figure}

TABLE \ref{errorresults} shows that as bandwidth doubles, the average error with removed outliers decreases from 79.06 m to 55.94 m, and then to 48.88 m. Fig.~\ref{fig:bandwidthcdfs} compares the CDFs of error and CRLBs for flights at a constant altitude with varying bandwidths. The lowest bandwidth (1.25 MHz) has the highest errors up to 200 m, while the highest bandwidth (5 MHz) performs best for the first 60\% of the data. In the final 40\%, the medium bandwidth (2.5 MHz) outperforms the rest. The CRLBs confirm that higher bandwidths reduce localization error. Increasing bandwidth improves TDOA localization by enhancing temporal resolution, allowing more precise measurement of signal arrival times. As shown in Equation \ref{eq:toa_crlb}, the variance of the TOA CRLB is inversely proportional to the square of the bandwidth, which reduces localization error \cite{gezici2008survey}. This increased precision reduces ambiguity, noise impact, and multipath effects, leading to more reliable distance estimates.

\subsection{Impact of NLOS on Localization Performance}

\begin{figure*}[htbp] 
    \centering
    \begin{subfigure}[b]{0.298\textwidth} 
        \centering
        \includegraphics[width=\textwidth, keepaspectratio]{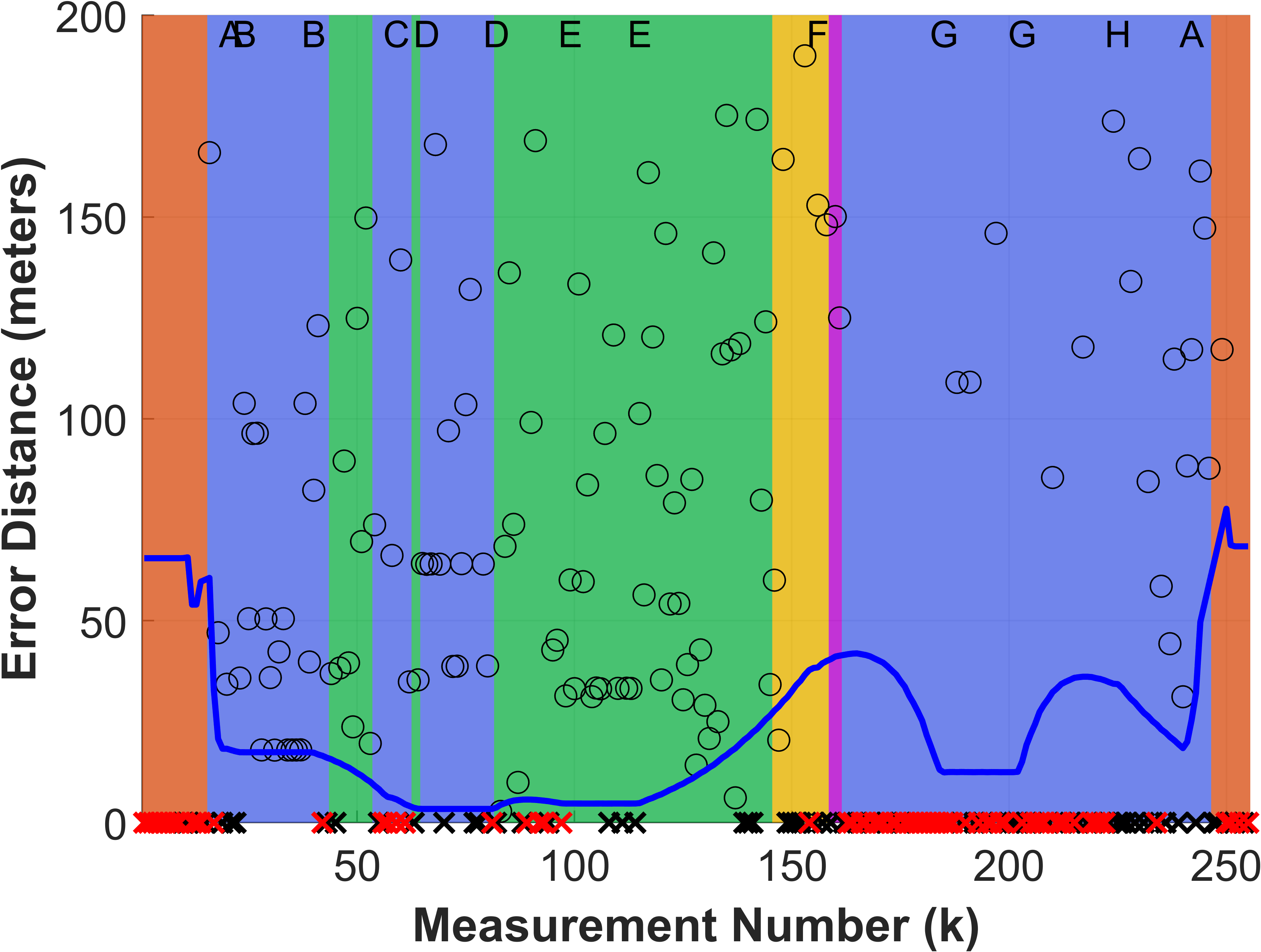} 
        \caption{40 m altitude, 1.25 MHz bandwidth.}
        \label{error40m1.25MHz}
    \end{subfigure}
    \hfill
    \begin{subfigure}[b]{0.298\textwidth} 
        \centering
        \includegraphics[width=\textwidth, keepaspectratio]{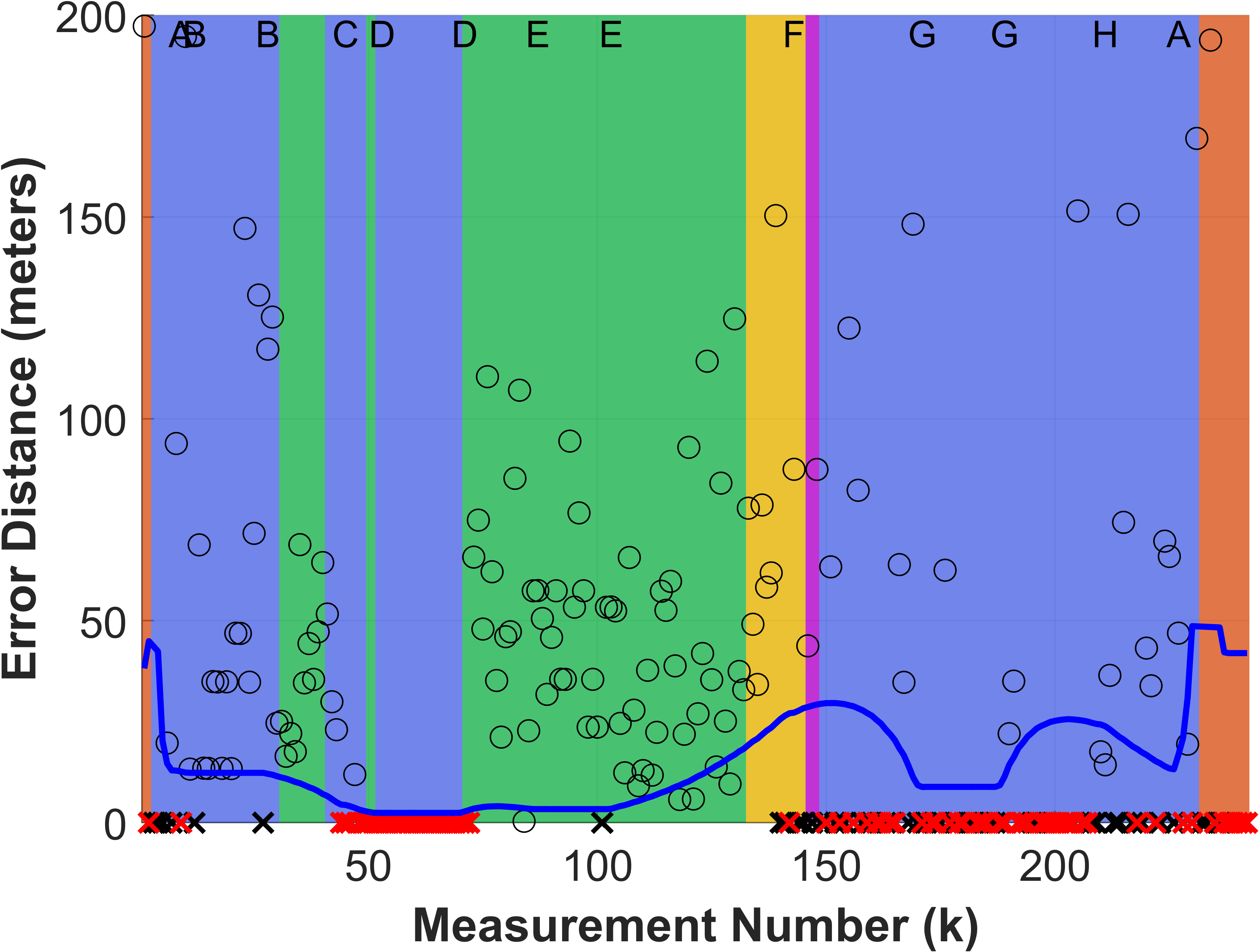} 
        \caption{40 m altitude, 2.5 MHz bandwidth.}
        \label{error40m2.5MHz}
    \end{subfigure}
    \hfill
    \begin{subfigure}[b]{0.298\textwidth} 
        \centering
        \includegraphics[width=\textwidth, keepaspectratio]{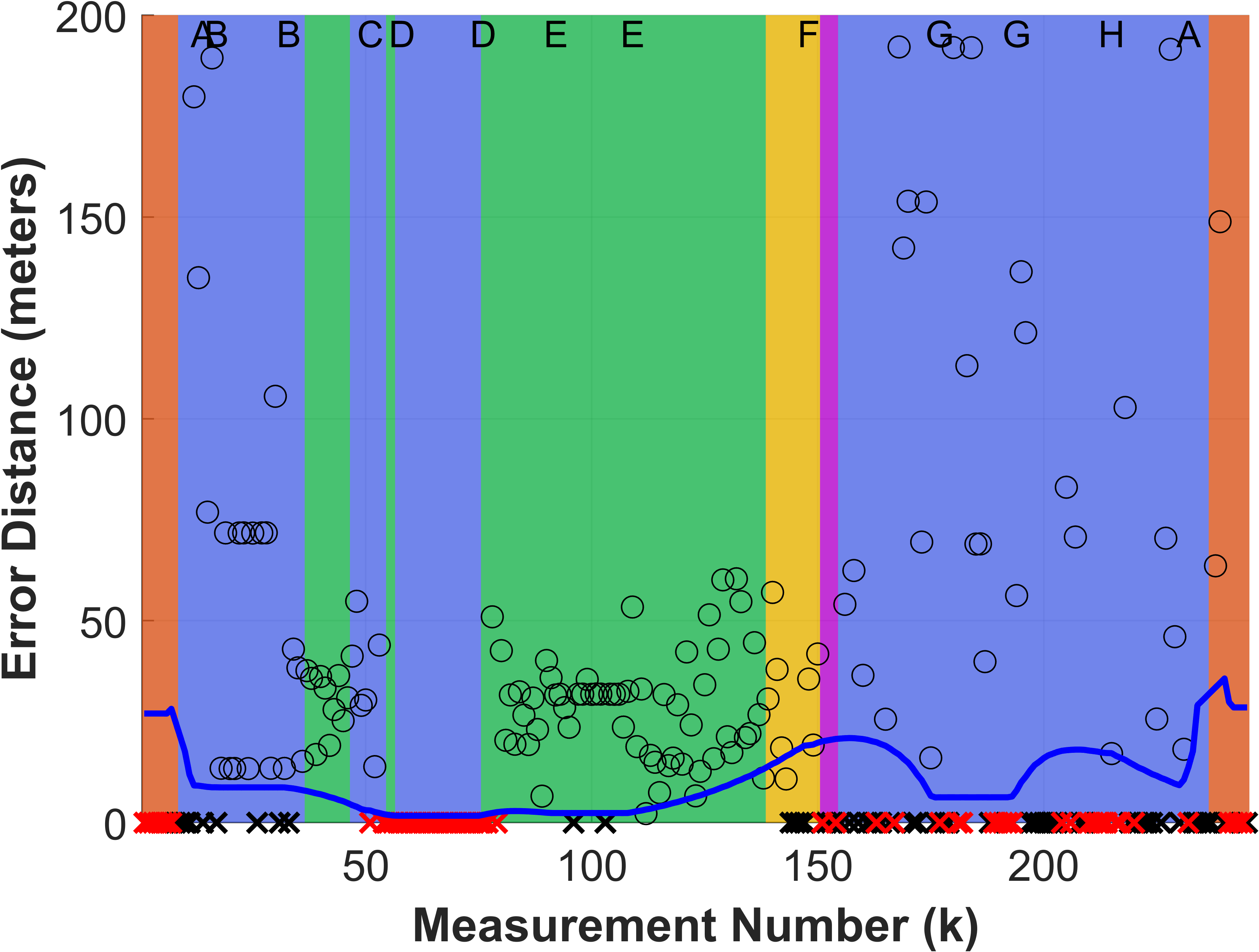} 
        \caption{40 m altitude, 5 MHz bandwidth.}
        \label{error40m5MHz}
    \end{subfigure}

    \vspace{0.5cm} 

    \begin{subfigure}[b]{0.298\textwidth} 
        \centering
        \includegraphics[width=\textwidth, keepaspectratio]{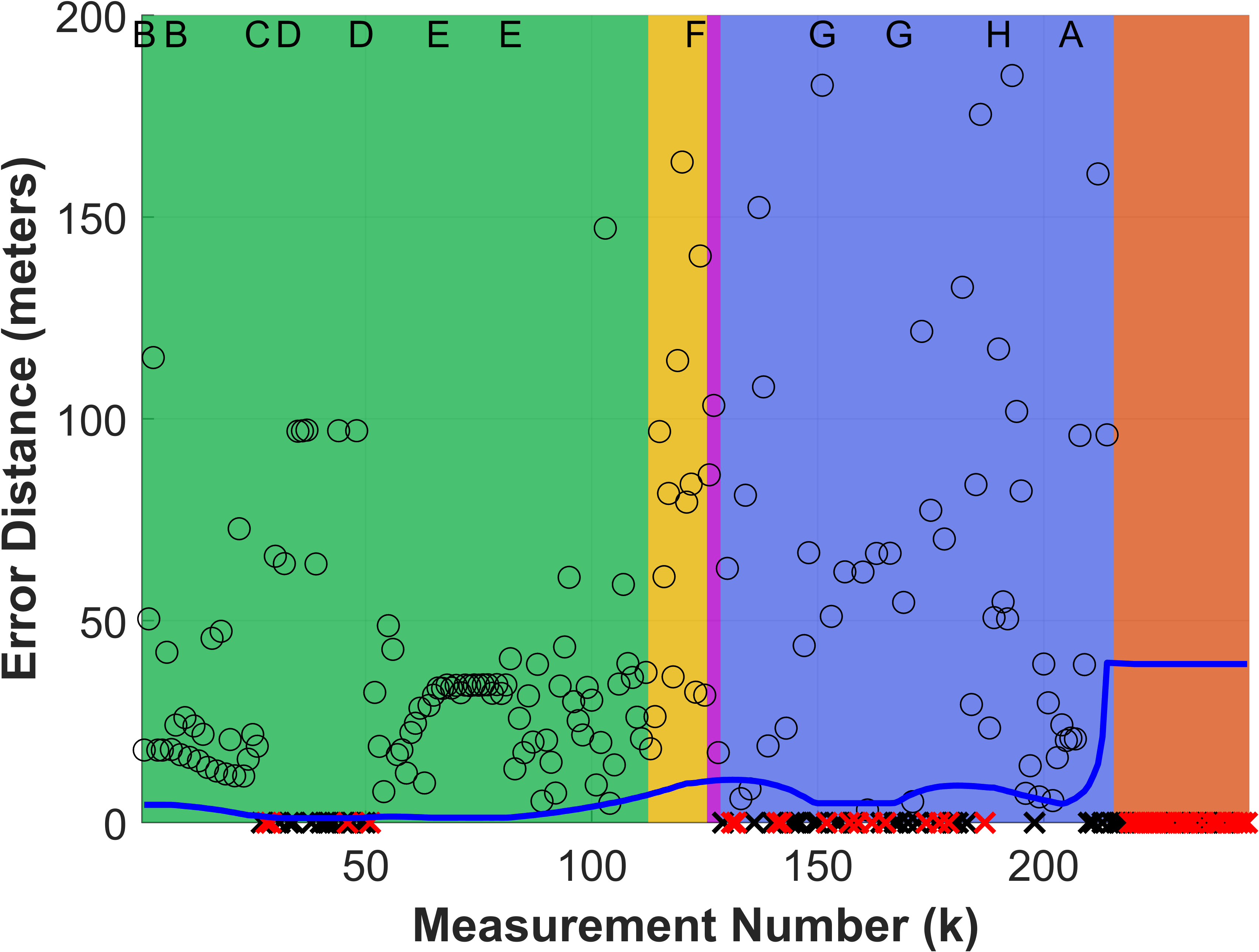} 
        \caption{70 m altitude, 5 MHz bandwidth.}
        \label{error70m5MHz}
    \end{subfigure}
    \hfill
    \begin{subfigure}[b]{0.298\textwidth} 
        \centering
        \includegraphics[width=\textwidth, keepaspectratio]{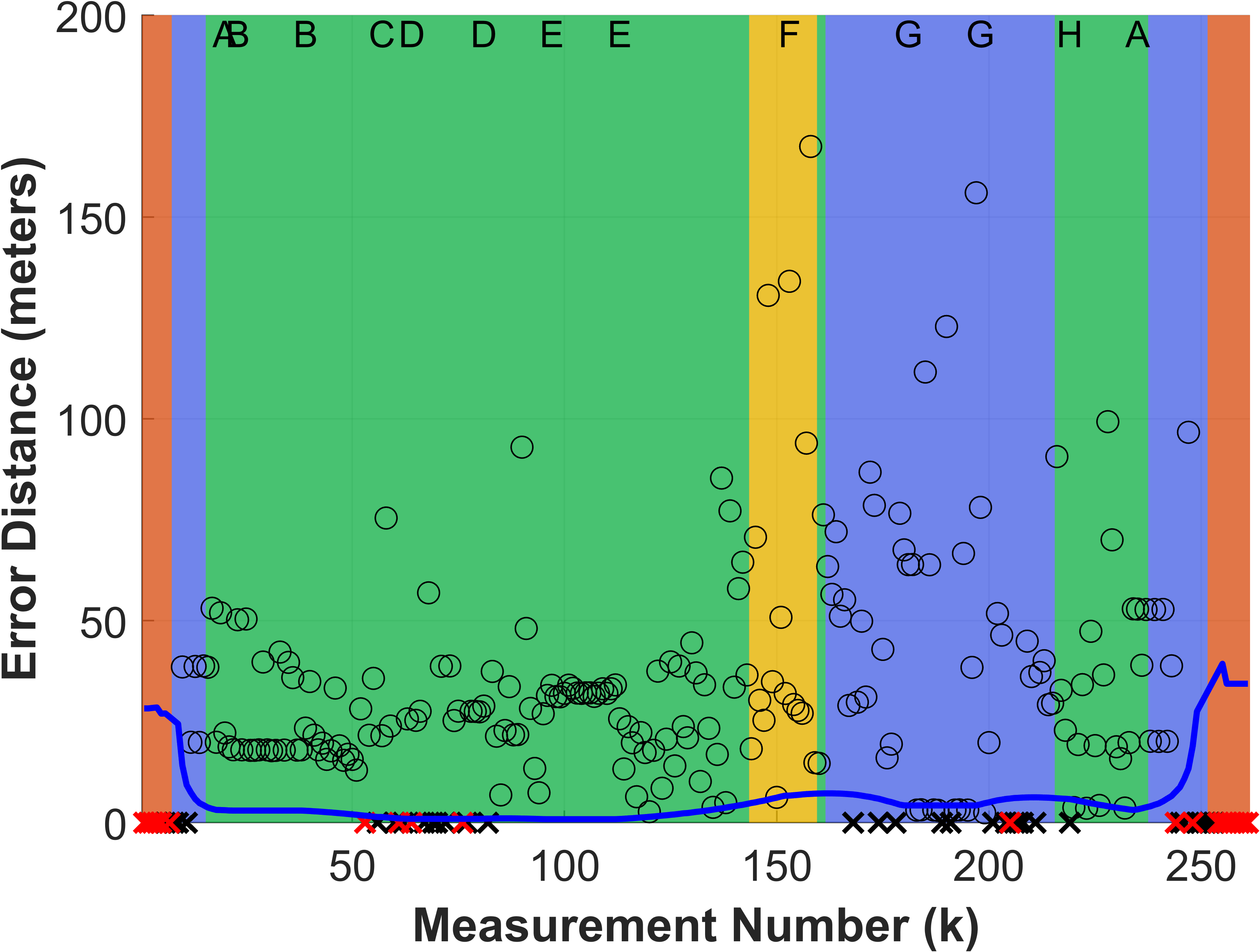} 
        \caption{100 m altitude, 5 MHz bandwidth.}
        \label{error100m5MHz}
    \end{subfigure}
    \hfill
    \begin{subfigure}[b]{0.298\textwidth} 
        \centering
        \scalebox{0.95}{ 
        \includegraphics[width=\textwidth, keepaspectratio]{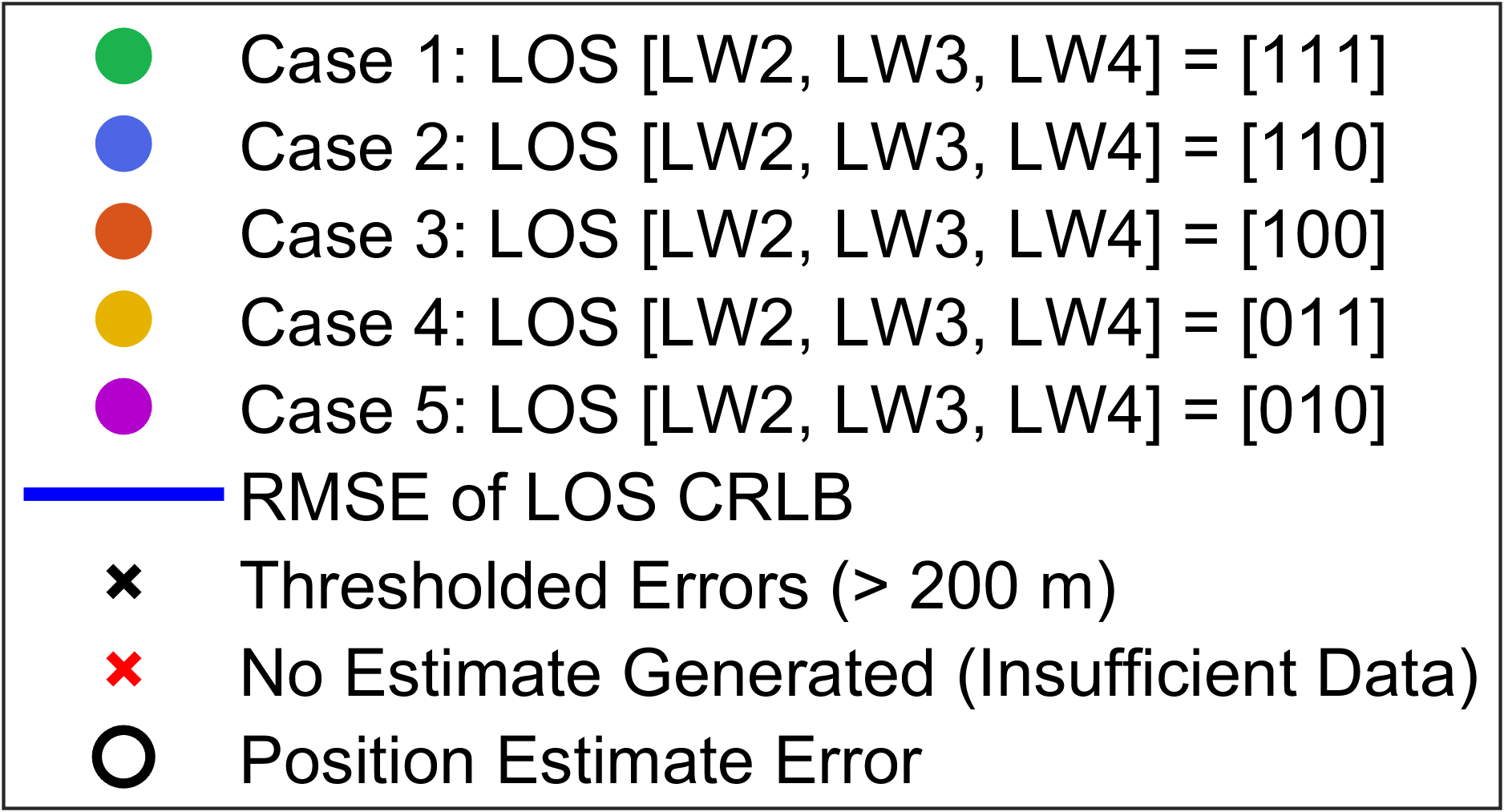}
        }
    \vspace{0.4cm} 
    \caption*{} 
    \end{subfigure}

    \caption{Error plots for each UAV flight, segmented by LOS/NLOS conditions for each tower. Position errors, CRLBs, thresholded measurements ($>$200 m), and missed estimates are shown, with waypoints labeled at the top of each subfigure (see also Fig.~\ref{fig:LOSconditions}).}
    \label{fig:LOSconditionsAndErrors}
\end{figure*}

Fig.~\ref{fig:LOSconditionsAndErrors} displays localization errors for each UAV flight, with black circles representing position errors (y-axis) and the CRLB shown as a blue line. The plots are segmented by LOS conditions, with waypoints from Fig. \ref{fig:LOSconditions} labeled at the top. Waypoints B, D, E, and G appear twice due to one-minute hovers at each of those points, while A, C, F, and H were passed without pauses. Red X's indicate points where no estimates were generated caused by poor geometry or NLOS blockages, while black X's mark errors over 200 meters. LW5 is excluded from LOS cases, as it was always NLOS. The figure highlights the influence of LOS conditions on localization performance. 

Fig. \ref{fig:LOSconditionsAndErrors} highlights the significant impact of NLOS obstructions on localization performance, introducing systemic biases and increased error variability. These obstructions degrade signals through multipath, attenuation, and reflection, leading to inaccuracies in TDOA-based localization. Performance was notably better when 3 of 4 sensors had LOS (Case 1) compared to scenarios with only 1 sensor in LOS (Cases 3 and 5). Error severity depends on both the number of obstructed sensors and their geometric importance, with obstructions to key sensors causing greater degradation than those in less critical positions. 
This underscores the importance of maintaining reliable timing references for accurate localization. While the RMSE of the LOS CRLB provides a theoretical benchmark for ideal LOS performance, localization errors in segments with two or more NLOS blockages often deviate significantly, highlighting the limitations of LOS-based models in NLOS conditions. Additionally, hovering directly over towers (Waypoint D) increases localization error despite LOS with 3 of 4 towers. This is likely due to the UAV's steep elevation angles relative to the RF sensors, where antenna gain decreases as elevation increases, reducing signal strength and SNR \cite{8690726}. Multipath effects and limited antenna sensitity at these angles further degrade performance. This underscores the need to account for 3D antenna radiation patterns in UAV localization, particularly near ground-based sensors.

\section{Conclusion}
This paper analyzed UAV localization accuracy using TDOA measurements under mixed LOS/NLOS conditions with real-world data. By comparing results to a 3D LOS TDOA CRLB model, we highlighted the impact of altitude and bandwidth on performance without NLOS bias. Higher bandwidths improved time resolution, while higher altitudes reduced NLOS biases and multipath effects, enhancing accuracy. Ray-tracing simulations enabled precise LOS/NLOS classification, improving the interpretability of results. The findings reveal the limitations of traditional LOS-based models in NLOS environments and highlight the need for adaptive models. Future work will incorporate NLOS biases into the CRLB model to address complex environmental factors, advancing reliable UAV localization for urban air traffic management and beyond.



\bibliographystyle{IEEEtran}
\bibliography{references.bib}

\end{document}